\documentclass[fleqn,twoside,twocolumn,nofootinbib,showkeys]{revtex4} 
\usepackage[sec,nocpr]{ujp} 

\begin{document}
\title[Magnetic Dynamics of a Multiferroic]
{MAGNETIC DYNAMICS OF A MULTIFERROIC\\ WITH AN ANTIFERROMAGNETIC LAYER}%
\author{S.V. Kondovych}
\affiliation{National Technical University of Ukraine
\textquotedblleft Kyiv
Polytechnical Institute\textquotedblright}
\address{37, Peremogy Ave., Kyiv 03056, Ukraine}
\email{ksvitlana@i.ua}
\author{H.V.~Gomonay}
\affiliation{National Technical University of Ukraine
\textquotedblleft Kyiv
Polytechnical Institute\textquotedblright}
\address{37, Peremogy Ave., Kyiv 03056, Ukraine}
\email{ksvitlana@i.ua}
\author{V.M.~Loktev}
\affiliation{Bogolyubov Institute for Theoretical Physics, Nat. Acad. of Sci. of Ukraine}
\address{14b, Metrolohichna Str., Kyiv 03680, Ukraine}


\udk{538.955+537.622.5+\\[-3pt] 537.226.86+537.635}
\pacs{75.85.+t,
75.70.-i,\\[-3pt]  75.50.Ee, 77.65.-j, 76.50.+g}
\razd{\secviii}

\autorcol{S.V.\hspace*{0.7mm}Kondovych, H.V.\hspace*{0.7mm}Gomonay,
V.M.\hspace*{0.7mm}Loktev}


\setcounter{page}{586}%

\begin{abstract}
Shape effects in magnetic particles are widely studied, because of the
ability of the shape and the size to control the parameters of a sample
during its production. Experiments with nano-sized samples show that
the shape can affect also the properties of antiferromagnetic (AFM)
materials. However, the theoretical interpretation of these effects
is under discussion. A model to study the shape-induced effects in AFM particles at the AFM
resonance frequency is proposed. The Lagrange function method is used to calculate
the spectrum of resonance oscillations of the AFM vector in a synthetic multiferroic
(piezoelectric + antiferromagnet). The influence of the specimen shape on the
AFM resonance frequency in the presence of an
external magnetic field is studied. Conditions for a resonance under the action
of an external force or for a parametric resonance to arise in the magnetic
subsystem are considered.
\end{abstract}

\maketitle

\section{Introduction}

Although the influence of the shape of ferromagnetic nanoparticles
on their magnetic properties has been widely studied
\cite{FM-1,FM-2,FM-3}, the shape effects for nano-dimensional
AFM particles remain a matter of discussion. In
works \cite{Folven1,Folven2,Folven3}, the effect of AFM vector
reorientation from a direction corresponding to the crystallographic
\textquotedblleft easy\textquotedblright\ axis to a direction
induced by a prolate particle was experimentally observed. This fact
enables us to talk about shape effects in antiferromagnets.

The majority of available methods used to observe the domain
structure in AFM materials and determine the AFM
vector orientation ~-- such as the X-ray polarization spectroscopy, the second
harmonic generation, and so forth~-- are intended to research static
effects. Often, they do not take the influence of the
dimensions and the shape of a specimen on its magnetic properties into account. We
suggest that a resonance method should be used for studying the
dynamics of the AFM vector. We will show that the shape-induced magnetic
anisotropy in AFM crystals can be analyzed with the help of the AFM
resonance spectrum or a spectrum of one of the long-wave quasiparticle
excitations. A significant contribution to the theory of the latter was
made by the outstanding Soviet and Ukrainian scientist O.S.~Davydov
\cite{Davydov}. This work is dedicated to the centennial anniversary
of his \mbox{birthday.}

More specifically, the work is aimed at studying a synthetic multiferroic
that combines AFM and piezoelectric (PE) properties. The purposes of the
work are (i)~to study the influence of time-dependent (periodic)
deformations of a piezoelectric subsystem on the magnetic subsystem in the
AFM/PE multiferroic, (ii)~to determine conditions, under which forced
resonance oscillations or a parametric resonance can arise in the
specimen, and (iii)~to analyze the dependence of resonance frequencies on
the geometrical parameters of the system in the presence of an external
magnetic field.

The main idea of the work consists in that we affect the elastic
subsystem of an AFM nanoparticle in order to study the dynamics of
its magnetic subsystem~-- namely, long-wave oscillations of the AFM
vector~-- and to analyze the contribution of the shape-induced
magnetic anisotropy. In so doing, we suppose that the mechanism of
influence of the shape on the magnetic anisotropy of AFM
nanoparticles has a magnetoelastic character, being governed by the
so-called \textquotedblleft destressing energy\textquotedblright\
\cite{GL2007}.

Note that the influence of the specimen shape on the AFM vector
dynamics was studied earlier. In particular, in work
\cite{Kornienko}, the dependence of the gap, which is a result
of the shape-induced anisotropy, in the AFM resonance
spectrum on the direction of an applied external magnetic field
has been analyzed. In this
work, a more complicated situation is considered, because the
proposed approach makes allowance for both the magnetic and elastic
subsystems in the AFM specimen. The perturbations in the
magnetoelastic system are simulated as those induced by an ac
electric voltage applied across a synthetic AFM/PE multiferroic. Besides
the specimen shape, the external electric and magnetic fields play the
role of control \mbox{parameters.}

\section{Model and Method}

\subsection{Researched specimen}

Consider a two-layer synthetic material fabricated by sputtering an
AFM film onto a PE substrate (Fig.~1,~\textit{a}). To simplify the
model, we consider that the antiferromagnet is collinear, including two
equivalent sublattices characterized by the magnetic moments
${\mathbf{M}}_{1}$ and ${\mathbf{M}}_{2}$. Let the specimen have an
elliptic shape with the semiaxis lengths $a$ and $b$, and let the AFM
film thickness be equal to $h$, with $a>b\gg h$. The dimensions of
the AFM subsystem are supposed to be small enough, so that it can be
considered in the \textquotedblleft macrospin\textquotedblright\
approximation, i.e. the dynamics of a uniform single-domain layer
can be analyzed. At temperatures below the N\'{e}el one, the
magnetic state of this system can be completely described with the use
of the AFM vector ${\mathbf{L}}={\mathbf{M}}_{1}-{\mathbf{M}}_{2}$
with a fixed length $\left\vert {\mathbf{L}}\right\vert =2M_{0}$.


\begin{figure}%
\vskip1mm
\includegraphics[width=\column]{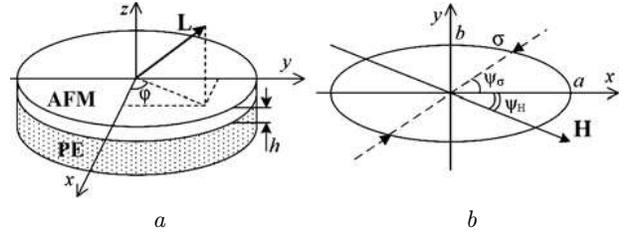}\\
{\it a\hspace{4cm}b} \vskip-3mm\caption{AFM film with thickness $h$
sputtered onto the PE substrate (\textit{a}). The AFM vector
$\mathbf{L}$ (the thick arrow) characterizes the magnetic state of
the system. Specimen has the elliptic shape with the semiaxis
lengths $a$ and $b$ (\textit{b}). Dashed arrows denote the direction
(the angle $\psi_{\sigma}$) of the force that creates a mechanical
stress in the specimen. The direction of strength lines of the
external magnetic field $\mathbf{H}$ is shown by the thin arrow, the
corresponding angle is ${\psi_{\mathrm{H}}}$ }\label{ellipses}
\end{figure}


Under the action of an external electric voltage $V$, there emerges a
mechanical stress in the piezoelectric, described by the stress
tensor $\sigma\sim Vd$, where $d$ is the piezoelectric tensor. This
stress brings about a specimen deformation
$u_{0}=\sigma_{0}/(2\mu)$, where $\mu$ is the shear modulus, and
$\sigma_{0}$ the component of the stress tensor in the direction of
the external force (at the angle $\psi_{\sigma}$ with respect to the
$Ox$ axis, see Fig.~1,~\textit{b}). The external magnetic field
${\mathbf{H}}=H\left(
\cos\psi_{\mathrm{H}},\sin\psi_{\mathrm{H}},0\right) $ lies in the
$xOy$ plane (Fig.~1,~\textit{b}).

\subsection{Lagrangian}

In order to calculate the AFM resonance spectrum, let us use a standard
technique based on the Lagrange function method (e.g., see
\cite{Turov}). The Lagrangian density for the AFM specimen equals
\[
{\mathcal L}_{\mathrm{AFM}} = \frac{\chi}{2g^2M_0^2
}{\mathrm{\dot{\mathbf L}}}^2 -
\frac{\chi}{2gM_0^2}(\dot{\mathbf{L}},\mathbf{L},\mathbf{H})\,+
\]\vspace*{-5mm}
\begin{equation}
\label{Lagrangian_AFM}
+\frac{\chi}{8M_0^2}[\mathbf{L}\times\mathbf{H}]^2 - w_{\mathrm{an}}
(\mathbf{L})-w_{\mathrm{destr}} - \Delta w_{\mathrm{destr}}
(\sigma),
\end{equation}
where $g$ is the gyromagnetic ratio; $\chi $ the magnetic susceptibility of
the material;
\begin{equation}
w_{\mathrm{an}}(\mathbf{L})=\frac{K_{\parallel
}}{4M_{0}^{2}}L_{z}^{2}-\frac{K_{\perp }}{16M_{0}^{4}}(L_{x}^{4}+L_{y}^{4})  \label{anisotropy_energy}
\end{equation}
is the density of the magnetic anisotropy energy for an AFM of the
\textquotedblleft easy plane\textquotedblright\
type\,\footnote{Anisotropy energy (\ref{anisotropy_energy})
corresponds to the tetragonal symmetry of a magnetic ordering
in the AFM layer.}; $K_{\parallel }\gg K_{\perp }>0$ are the
anisotropy constants;
\[
w_{\mathrm{destr}}=\frac{1}{2}\biggl\{\!\frac{K^{\mathrm{el}}_{\parallel}}{4M_0^2}
(L_y^2-L_x^2)+
  \frac{K^{\mathrm{el}}_{\mathrm{is}}}{16M_0^4}\biggl[\left(
  L_x^2-L_y^2\right)^2 +
  \]\vspace*{-5mm}
\begin{equation}\label{destressing_energy}
+ \, 4\left( L_xL_y\right)^2\biggr]-
  \frac{K^{\mathrm{el}}_{\perp}}{16M_0^4}\left[\!\left(
  L_x^2-L_y^2\right)^2-4\left(
L_xL_y\right)^2\!\right]\!\biggr\}
\end{equation}
is the destressing energy density\,\footnote{Destressing energy
(\ref{destressing_energy}) is written down for a single-domain film
(i.e. the geometrical dimensions of the specimen are supposed to be
small enough, so that a non-uniformity of the magnetic subsystem can
be neglected) and assuming that the elastic properties of the
crystal are isotropic.} \cite{GL2007}; $K_{\parallel
}^{\mathrm{el}}$, $K_{\mathrm{is}}^{\mathrm{el}},$ and $K_{\perp
}^{\mathrm{el}}$ are the magnetoelastic coefficients depending on
the specimen shape (see Appendix); and\vspace*{-1mm}
\[
\Delta w_{\mathrm{destr}} (\sigma) = \frac{1}{4\mu}\sigma_0 \cos
2\psi_{\sigma} \biggl\{\!(K^{\mathrm{el}}_{\parallel}+\Delta
K_{\parallel})\frac{L_y^2-L_x^2}{4M_0^2}\, -  \]\vspace*{-5mm}
\begin{equation}\label{d_w_destr}
 - \left(K^{\mathrm{el}}_{\perp}+\Delta K_{\perp}\right) \frac{\left(
  L_x^2-L_y^2\right)^2-4\left(
L_xL_y\right)^2 }{16M_0^4} \!\biggr\}
\end{equation}
is a correction to $w_{\mathrm{destr}}$ that arises under the
influence of an external stress owing to a variation of the specimen
shape at a deformation \cite{lviv} (expressions for $\Delta
K_{\parallel ,\perp }$ can be found in Appendix).\vspace*{-1mm}

\subsection{Parametrization of the AFM vector}

Let us examine the emergence of resonance oscillations of the AFM
vector $\mathbf{L}$ in a vicinity of its equilibrium position
${\mathbf{L}}_{\mathrm{eq}}$ and analyze the influence of the
specimen shape on its resonance frequencies. In the absence of
external fields, the AFM vector lies in the $xOy$ plane, and it can
be parametrized by the angle $\varphi _{\mathrm{eq}}$ with respect
to the $Ox$ axis: ${\mathbf{L}}_{\mathrm{eq}}=$\linebreak
$=2M_{0}\left( \cos \varphi _{\mathrm{eq}},\sin \varphi
_{\mathrm{eq}},0\right) $. If the crystallographic axes in the
material are directed along the ellipse axes, and the magnetoelastic
contribution (Eqs.~(\ref{destressing_energy}) and (\ref{d_w_destr}))
is not taken into account, this angle can be equal to $\varphi
_{\mathrm{eq1}}=0$ or $\varphi _{\mathrm{eq2}}=\pi /2$; in the
general case, the angle $\varphi _{\mathrm{eq}}$ can accept an
arbitrary value. The external magnetic field also affects the
orientation of \mbox{AFM vector.}\looseness=-1

Let the structures of the crystal lattices in the AFM and PE substances
coincide, and let the substrate deformation invoked by an external
electric voltage be transferred into the AFM film, which results in
a small rotation of the AFM vector,
${\mathbf{L}}={\mathbf{L}}_{\mathrm{eq}}+{\mathbf{l}}$, where
$\left\vert {\mathbf{l}}\right\vert \ll \left\vert
{\mathbf{L}}_{\mathrm{eq}}\right\vert $ and ${\mathbf{l}}\perp
{\mathbf{L}}_{\mathrm{eq}}$. Then it is convenient to introduce the
following parametrization for the \mbox{vector $\mathbf{L}$:}
\[
L_x = 2M_0 \cos\varphi_{\mathrm{eq}} ( 1 - l_{\perp}^2/2 -
l_{\parallel}^2/2 ) - 2M_0 l_{\perp} \sin\varphi_{\mathrm{eq}};
\]\vspace*{-6mm}
\[
L_y = 2M_0 \sin\varphi_{\mathrm{eq}} ( 1 - l_{\perp}^2/2 -
l_{\parallel}^2/2 ) + 2M_0 l_{\perp} \cos\varphi_{\mathrm{eq}};
\]\vspace*{-6mm}
\begin{equation} \label{param}
L_z = 2M_0 l_{\parallel}.
\end{equation}

Here, $l_{\perp }$ and $l_{\parallel }$ are the projections of the vector
$\mathbf{l}$ on the plane and the $Oz$ axis, respectively.

\section{Oscillations of the AFM Vector\\ under the Action of an External Force}

Let the AFM/PE specimen be unstrained in the absence of external fields. We
express the alternating mechanical stress that arises in the specimen under
the action of an external electric voltage with frequency $\omega $ in the
form $\sigma (t)=\mathrm{Re}\left\{ \sigma _{0}e^{i\omega t}\right\} $ and
suppose the amplitude $\sigma _{0}$ to be small. Being determined from the
standard procedure of potential energy minimization, the stable equilibrium
states of the AFM vector (they are parametrized by the angles $\varphi
_{\mathrm{eq}}$) satisfy the relations
\[
\left( 2K_{\perp} + 4K^{\mathrm{el*}}_{\perp} \right) \sin
4\varphi_{\mathrm{eq}} + 2K^{\mathrm{el*}}_{\parallel} \sin
2\varphi_{\mathrm{eq}} \,-
\]\vspace*{-7mm}
\begin{equation} \label{eq_magn}
 - \,\chi H^2
\sin 2(\varphi_{\mathrm{eq}}-\psi_{\mathrm{H}})=0;
\end{equation}\vspace*{-7mm}
\[
\Omega_{\perp}^2 \equiv \frac{g^2}{2\chi} \biggl[ \left( 2K_{\perp}
+ 4K^{\mathrm{el*}}_{\perp} \right) \cos 4\varphi_{\mathrm{eq}} +
K^{\mathrm{el*}}_{\parallel} \cos 2\varphi_{\mathrm{eq}} \,-
\]\vspace*{-7mm}
\begin{equation} \label{stable}
- \frac{1}{2}\chi {\mathrm H}^2 \cos
2(\varphi_{\mathrm{eq}}-\psi_{\mathrm H}) \biggr] \ge 0; \quad
\Omega_{\parallel}^2 \approx \frac{g^2 K_{\parallel}}{2\chi} \ge 0,
\end{equation}
where the quantities $\Omega _{\perp }$ and $\Omega _{\parallel }$ coincide
with the frequencies of characteristic free oscillations of the AFM vector in
the specimen plane and perpendicularly to it, respectively (see below). The
magnetoelastic coefficients $K_{\parallel }^{\mathrm{el\ast }}$ and
$K_{\perp }^{\mathrm{el\ast }}$ include corrections $\Delta K_{\parallel
,\perp }$ that arise under the action of a mechanical stress $\sigma (t)$
applied in the $\psi _{\sigma }$-direction,
\begin{equation}\label{elas_add}
\begin{array}{l}
 \displaystyle K_{\parallel}^{\mathrm{el*}} = K^{\mathrm{el}}_{\parallel} +
\frac{1}{2\mu}\sigma (K^{\mathrm{el}}_{\parallel}+\Delta
K_{\parallel})\cos 2\psi_{\sigma} ; \\[5mm]
 \displaystyle K_{\perp}^{\mathrm{el*}} = K^{\mathrm{el}}_{\perp} +
\frac{1}{2\mu}\sigma \left(K^{\mathrm{el}}_{\perp}+\Delta
K_{\perp}\right) \cos 2\psi_{\sigma}.
\end{array}
\end{equation}

In order to obtain the equations describing the AFM vector oscillations, let
us parametrize Lagrangian (\ref{Lagrangian_AFM}) and pass to small
deviations $l_{\perp }$ and $l_{\parallel }$ from the equilibrium positions
(see Eq.~(\ref{param})). Taking expression (\ref{eq_magn}) for
equilibrium states of the AFM vector into account and neglecting the terms proportional to
$\sigma _{0}l_{\perp }^{2}$ and $\sigma _{0}l_{\parallel }^{2}$ as the
quantities of the third order of smallness, we obtain the Lagrange
equations
\begin{equation}\label{Lagr_equat}
\begin{array}{l}
 \displaystyle
\ddot l_{\perp} + 2\gamma_{\mathrm{AFM}} \dot l_{\perp}
-\Omega_{\mathrm H} \dot l_{\parallel} + \Omega_{\perp}^2 l_{\perp}
= - \Phi \sigma_0 e^{i \omega t}; \\[3mm]
\displaystyle \ddot l_{\parallel} + 2\gamma_{\mathrm{AFM}} \dot
l_{\parallel} + \Omega_{\mathrm H} \dot l_{\perp} +
\Omega_{\parallel}^2 l_{\parallel} = 0.
\end{array}
\end{equation}
The coefficient $\gamma _{\mathrm{AFM}}$ is the AFM resonance width;
it simulates the damping of oscillations. In
Eqs.~(\ref{Lagr_equat}), the following notations wwere introduced:
\[
\Omega_{\mathrm H} = g{\mathrm H} \cos
(\varphi_{\mathrm{eq}}-\psi_{\mathrm H});
\]
\[
\Phi \equiv \Phi \left( \Delta K_{\parallel,\perp},
\varphi_{\mathrm{eq}}, \psi_{\sigma} \right) = \frac{1}{8\mu}
\frac{g^2}{\chi} \biggl[ (K^{\mathrm{el*}}_{\parallel}+\Delta
K_{\parallel}) \times
\]\vspace*{-7mm}
\[ \times \sin 2\varphi_{\mathrm{eq}} +
2\left(K^{\mathrm{el*}}_{\perp}+\Delta K_{\perp}\right)
 \sin 4\varphi_{\mathrm{eq}} \biggr] \cos
2\psi_{\sigma},
\]
$\varphi _{\mathrm{eq}}$ is the solution of Eq.~(\ref{eq_magn}), and
the angles $\psi _{\mathrm{H}}$ and $\psi _{\sigma }$ are defined
above (see Fig.~1,~\textit{b}). The function $\Phi $ in
Eq.~(\ref{Lagr_equat}) makes allowance for a variation of the specimen
shape at a deformation and modifies the magnitude of external
influence. By changing the angle $\psi _{\sigma }$ that determines
the specimen deformation direction, it is possible to control the
amplitude of AFM vector oscillations. The frequencies $\Omega
_{\perp }$ and $\Omega _{\parallel }$ are the frequencies of
characteristic oscillations of the AFM vector components $l_{\perp }$
and $l_{\parallel }$, respectively. They are different for different
equilibrium orientations of the vector $\mathbf{L}$ and depend on both
the magnetoelastic constants and the magnetic field $\mathbf{H}$.
Note that $\Omega _{\parallel }\gg \Omega _{\perp }$, because
$K_{\parallel }\gg K_{\perp }$.

The resonance in the AFM/PE system arises at the
frequencies\vspace*{-3mm}
\[
\omega^2 = \omega_{1,2}^2 =\frac 1 2 \biggl( \Omega_{\perp}^2 +
\Omega_{\parallel}^2 + \Omega_{\mathrm H}^2 -
2\gamma_{\mathrm{AFM}}^2 \,\pm  \]\vspace*{-7mm}
\begin{equation}\label{frequencies}
{\pm} \,\sqrt{\left( \Omega_{\perp}^2 + \Omega_{\parallel}^2 +
\Omega_{\mathrm H}^2 \right)^2 - 4 \Omega_{\perp}^2
\Omega_{\parallel}^2 - 4 \gamma_{\mathrm{AFM}}^2 \Omega_{\mathrm
H}^2 } \biggr).
\end{equation}

In the absence of a magnetic field, there exist four different equilibrium
orientations of the AFM vector in a single-domain specimen (they can be obtained
from Eq.~(\ref{eq_magn}) with ${\mathbf{H}}=0$),
\begin{equation}\label{states}
\begin{array}{l}
 \displaystyle
\varphi_{\mathrm{eq1}} = 0;
\quad\varphi_{\mathrm{eq2}}=\frac{\pi}{2};
\\[3mm]
 \displaystyle\cos 2\varphi_{\mathrm{eq3,4}} = \frac
{-K^{\mathrm{el*}}_{\parallel}}{(2 K_{\perp} + 4
K^{\mathrm{el*}}_{\perp})}.
\end{array}
\end{equation}

The stability of the equilibrium states (\ref{states}) was studied in
work \cite{lviv}. Note that no more than two orientations of the vector
${\mathbf{L}}$ can be stable simultaneously (see Fig.~2): either at
the angles $\varphi _{\mathrm{eq1,2}}$ or the angles
$\varphi_{\mathrm{eq3,4}}$. If the ellipse is prolate and the ratio
between the lengths of its semiaxes exceeds a certain critical value
$(a/b)_{\mathrm{crit}}$ \cite{GL2007}, there exists only one stable
equilibrium orientation for the AFM vector--along the $Ox$ axis.


\begin{figure}%
\vskip1mm
\includegraphics[width=\column]{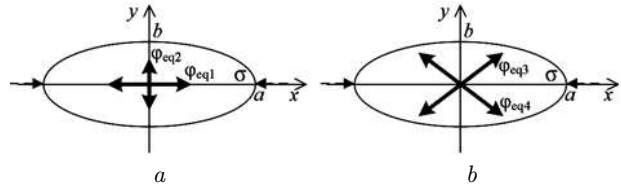}\\
{\it a\hspace{4cm}b} \vskip-3mm\caption{Equilibrium orientations of
the AFM vector. Dashed arrows denote the specimen deformation
direction (along the $Ox$ axis). Thick arrows correspond to the
states of the AFM vector that can be stable simultaneously:
(\textit{a})~at the angles $\varphi_{\mathrm{eq1}}=0$ and $\varphi
_{\mathrm{eq2}}=\pi/2$ and (\textit{b}) at the angles
$\varphi_{\mathrm{eq3,4}}$ (see Eq.~(\ref{states})) with respect to
the $Ox$ axis }\label{ellipses2}\vspace*{-2mm}
\end{figure}

%

For definiteness, let the specimen be deformed along the longer
ellipse axis ($\psi _{\sigma }=0$). The
de\-for\-ma\-ti\-on-in\-du\-ced modification of the specimen shape
affects the behavior of the magnetic system, and this influence
depends on the stable state, in which the AFM vector dwells. In a
vicinity of the equilibrium orientation at the angle $\varphi
_{\mathrm{eq1}}=0$ or $\varphi _{\mathrm{eq2}}=\pi /2$, there is no
resonance under the action of an external force, because $\Phi =0$
in the system of equations (\ref{Lagr_equat}). A deformation will
result in only a shift of characteristic oscillation frequencies
$\Omega _{\perp }$ and $\Omega _{\parallel }$ according to
corrections in Eq.~(\ref{elas_add}).

At the same time, the emergence of oscillations in the AFM/PE system
owing to the shape change is possible in a vicinity of the equilibrium
angles $\varphi _{\mathrm{eq3,4}}$ (Fig.~2,\textit{b}). The
effective projection of a mechanical stress $\sigma $ deflects the AFM
vector from its stable equilibrium orientation, and resonance
oscillations of the vector $\mathbf{L}$ in the specimen plane can occur
even in the absence of a magnetic field (Fig.~3).

It should be emphasized that, for the resonance to take place under
such conditions, the geometry of the system and the relative
directions between the axes of an elliptic specimen and the
crystallographic axes of material must be selected carefully to
provide the condition $K_{\perp }^{\mathrm{el}}<0$, which is
responsible for the stability of \textquotedblleft
diagonal\textquotedblright\ equilibrium positions. The analysis of
Eq.~(\ref{Lagr_equat}) with a non-zero magnetic field shows that the
oscillations of individual AFM vector components become
\textquotedblleft entangled\textquotedblright, so that the vector
$\mathbf{L}$ oscillates both in the specimen plane and
perpendicularly to it. Neglecting the oscillation damping
($\gamma_{\mathrm{AFM}}=0$) and writing down the corresponding
solutions for the components of the AFM vector in the
forms\vspace*{-2mm}
\begin{equation}
\begin{array}{l}
 \displaystyle
l_{\perp} =  \frac {\Phi \sigma_0 \left( \omega^2 -
\Omega_{\parallel}^2 \right)} {\left( \omega^2 - \Omega_{\perp}^2
\right)\left( \omega^2 - \Omega_{\parallel}^2 \right) -
\Omega_{\mathrm H}^2 \omega^2} \cos \omega t; \\[6mm]
 \displaystyle l_{\parallel} = -
\frac {\Phi \sigma_0 \Omega_{\mathrm H} \omega}{\left( \omega^2 -
\Omega_{\perp}^2 \right)\left( \omega^2 - \Omega_{\parallel}^2
\right) - \Omega_{\mathrm H}^2 \omega^2} \sin \omega t,
\end{array}
\end{equation}

\begin{figure}%
\vskip1mm
\includegraphics[width=5cm]{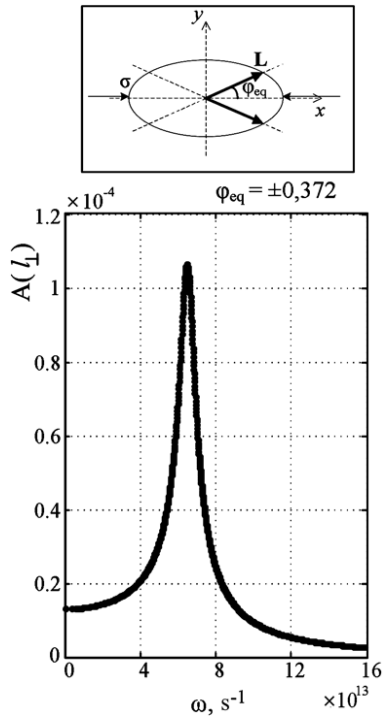}
\vskip-3mm\caption{Amplitude-frequency characteristic of
oscillations of the AFM vector $\mathbf{L}$. The external magnetic
field is absent. $A(l_{\perp})$ is the amplitude of a relative
deviation of the vector $\mathbf{L}$ from the equilibrium
orientation. The inset: the vector $\mathbf{L}$ (the thick arrow)
oscillates in the $xOy$ plane, the angles $\varphi_{\mathrm{eq}}$
correspond to the stable equilibrium orientations of the vector
$\mathbf{L}$, thin arrows denote the direction of an external
mechanical stress $\sigma$ (along the $Ox$ axis, $\psi_{\sigma}=0$).
Simulation was carried out for an NiO AFM particle with the ratio
between the ellipse semiaxis lengths $a/b=20$. The other model
parameters are
quoted in Table }\label{res1}
\end{figure}

\begin{table}[h!]
\vskip3mm \noindent\caption{Physical characteristics of NiO\\
antiferromagnet used in the simulation\\ (see works
\cite{GL2007,NiO-1,NiO-2,NiO-3} and
Appendix)}\vskip3mm\tabcolsep18.7pt

\noindent{\footnotesize\begin{tabular}{|l|c|c|}
  \hline
 \multicolumn{1}{|c}{Parameter} &
  \multicolumn{1}{|c}{$a/b = 2.4$ } &
   \multicolumn{1}{|c|}{\rule{0pt}{5mm}$a/b = 20$} \\[2mm]
  \hline
 \multicolumn{1}{|l}{\rule{0pt}{5mm}~~$\chi$} &\multicolumn{2}{|c|}{$1.28 \times
10^{-3}$}\\
~~$g$, s$^{-1}$/Т&\multicolumn{2}{|c|}{$2.5\gamma_e = 4.4 \times 10^{11}$}\\
~~$K_{\parallel}$, J/m$^3$&\multicolumn{2}{|c|}{$ 500$}\\%
~~$K_{\perp}$, J/m$^3$&\multicolumn{2}{|c|}{$28.8$}\\%
~~$K^{\mathrm{el}}_{\parallel}$, J/m$^3$&20&$90$\\%
~~$| K^{\mathrm{el}}_{\perp} |$, J/m$^3$&5&$45$\\[2mm]%
  \hline
\end{tabular}
 }\vskip-3mm
\end{table}

\begin{figure*}%
\vskip1mm
\includegraphics[width=13.3cm]{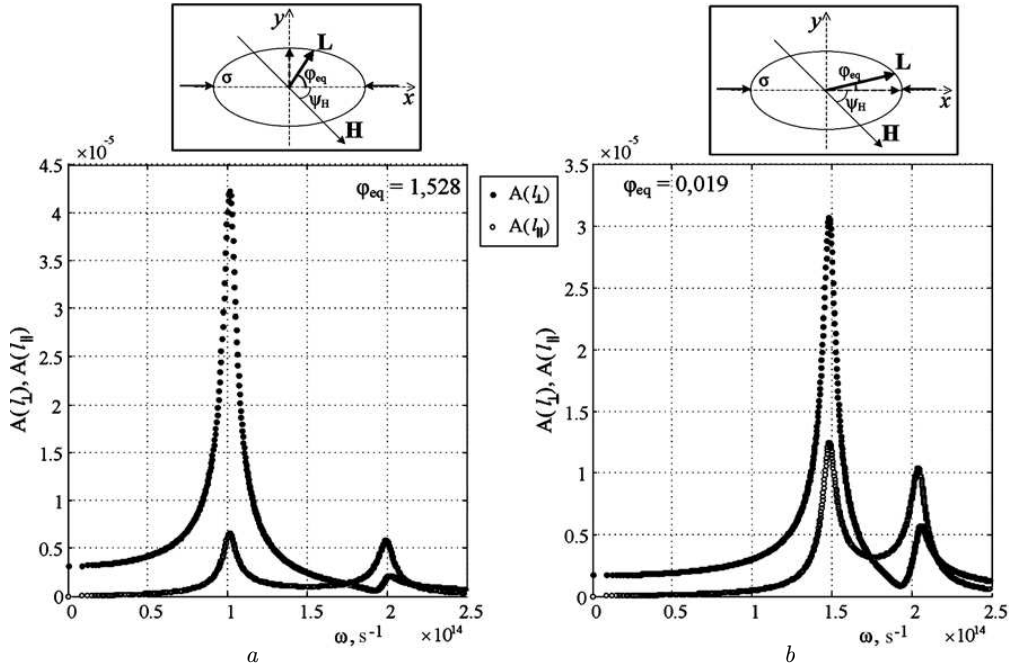}\\[-2mm]
{\it ~~a\hspace{7.0cm}b} \vskip-3mm\caption{Amplitude-frequency
characteristics of oscillations of the AFM vector $\mathbf{L}$ in a
magnetic field calculated from expression (\ref{frequencies}).
$A(l_{\perp})$ and $A(l_{\parallel})$ are the amplitudes of relative
deviations of the vector $\mathbf{L}$ from equilibrium orientations
in the $xOy$ plane and along the $Oz$ axis, respectively. Panels
(\textit{a}) and (\textit{b}) correspond to different equilibrium
orientations of the AFM vector (\ref{eq_magn}) schematically
indicated \textit{in the insets} by dashed arrows. Thin arrows
\textit{in the insets} denote the direction of an external
mechanical stress $\sigma$ (along the $Ox$ axis, $\psi_{\sigma}=0$).
Simulation was carried out for an NiO AFM particle with the ratio
between the ellipse semiaxis lengths $a/b=20$. The other model
parameters are quoted in Table. The external magnetic field
$\mathbf{H}$ is directed at the angle $\psi_{\mathrm{H}}=-\pi/4$.
$H=100~\mathrm{Oe}
\approx7.7\times10^{3}$~\textrm{A/m}\vspace*{-2mm} }\label{res2}
\end{figure*}

\noindent one can immediately see that the application of an
external magnetic field changes the character of oscillations.
Namely, instead of two linearly polarized modes (a \textquotedblleft
soft\textquotedblright\ mode in the $xOy $ plane and a
\textquotedblleft rigid\textquotedblright\ one perpendicularly to
it), the oscillations of the vector $\mathbf{L}$ are described by
elliptically polarized modes, whose ellipticity degree depends on
$\mathbf{H}$. Figure~4 demonstrates the difference of the
oscillation amplitudes for the components $l_{\perp}$ and
$l_{\parallel}$. The resonance frequencies of those modes depend on
the equilibrium AFM vector orientation, in a vicinity of which the
vector oscillates, and are given by \mbox{expression
(\ref{frequencies}).}\looseness=1

If ${\mathbf{H}}\neq0$, the equilibrium orientations of the AFM
vector change; namely, the magnetic field rotates the vector
$\mathbf{L}$ (see insets in Fig.~4). Let the equilibrium position of
the AFM vector in a vicinity of 0 be designated as
$\varphi_{\mathrm{eq1}}$ and near $\pi/2$ as $\varphi
_{\mathrm{eq2}}$; those angles are the solutions of
Eq.~(\ref{eq_magn}). Figure~4 exhibits the amplitude-frequency
characteristics of oscillations in a vicinity of the equilibrium
states $\varphi_{\mathrm{eq1}}$ and $\varphi _{\mathrm{eq2}}$. The
values of maximum positions (resonance frequencies
(\ref{frequencies})) are changed if the specimen shape is involved
in the Lagrangian (it is so because the corrections $\Delta
K_{\parallel,\perp}$ depend on the ratio $a/b$ between the ellipse
semiaxis lengths, which parametrizes the specimen shape). The
orientation of the AFM vector is determined not only by the
directions of the crystallographic axes and the direction of an
external magnetic field, but also by the direction of the
\textquotedblleft easy\textquotedblright\ axis given by the particle
shape (along the larger ellipse axis). The shape-induced uniaxial
magnetic anisotropy eliminates the energy degeneration among various
equilibrium states of the AFM vector (in the infinite-specimen
approximation and in the absence of external fields, the states
$\varphi=0$ and $\varphi=\pi/2$ are equivalent), and the frequencies
of oscillations near those equilibrium orientations \mbox{become
split.}\looseness=1

The dependences of the oscillation frequencies on the direction and
the magnitude of an external magnetic field, as well as on the
elliptic specimen shape, are depicted in Fig.~5. The figure
demonstrates a number of regularities:

1. The specimen shape, being parametrized by the ratio between the
ellipse semiaxis lengths $a/b$, affects the magnitudes of resonance
frequencies. When comparing cases $a$ and $c$ in Fig.~5, one can see
that the \textquotedblleft soft\textquotedblright\ mode frequency
$\omega_{2}$ becomes almost twice as large if the eccentricity $a/b$
increases. The \textquotedblleft rigid\textquotedblright\ mode
frequency $\omega_{1}$ remains almost constant at that. This fact
follows from that the deformation makes a larger correction to
$K_{\parallel}^{\mathrm{el}}$ than to $K_{\perp }^{\mathrm{el}}$
(\ref{elas_add}) as the parameter $a/b$ increases.

2. The \textquotedblleft soft\textquotedblright\ modes are more
sensitive to ani\-so\-tro\-py. It is clear because the dependences
on the magnetic field direction are stretched along different
directions for different equilibrium states. The axis, along which
the angular diagram is stretched, corresponds to the state with
lower energy (the mutually perpendicular orientations of the vectors
$\mathbf{H}$ and $\mathbf{L}$).

3. If the parameter $a/b$ grows, the frequency $\omega_{2}$ of AFM
vector oscillations in a vicinity of
$\varphi_{\mathrm{eq2}}=\pi/2$ decreases at certain magnetic field
directions (the angle $\psi_{\mathrm{H}}$). This is evidenced by a
bending of the angular diagrams in Figs.~5,~\textit{b} and \textit{d}.
The obtained dependence allows the influence of the AFM/PE specimen
shape on the dynamics of its magnetic subsystem to be
examined.\vspace*{-2mm}

\begin{figure*}
\includegraphics[width=15cm]{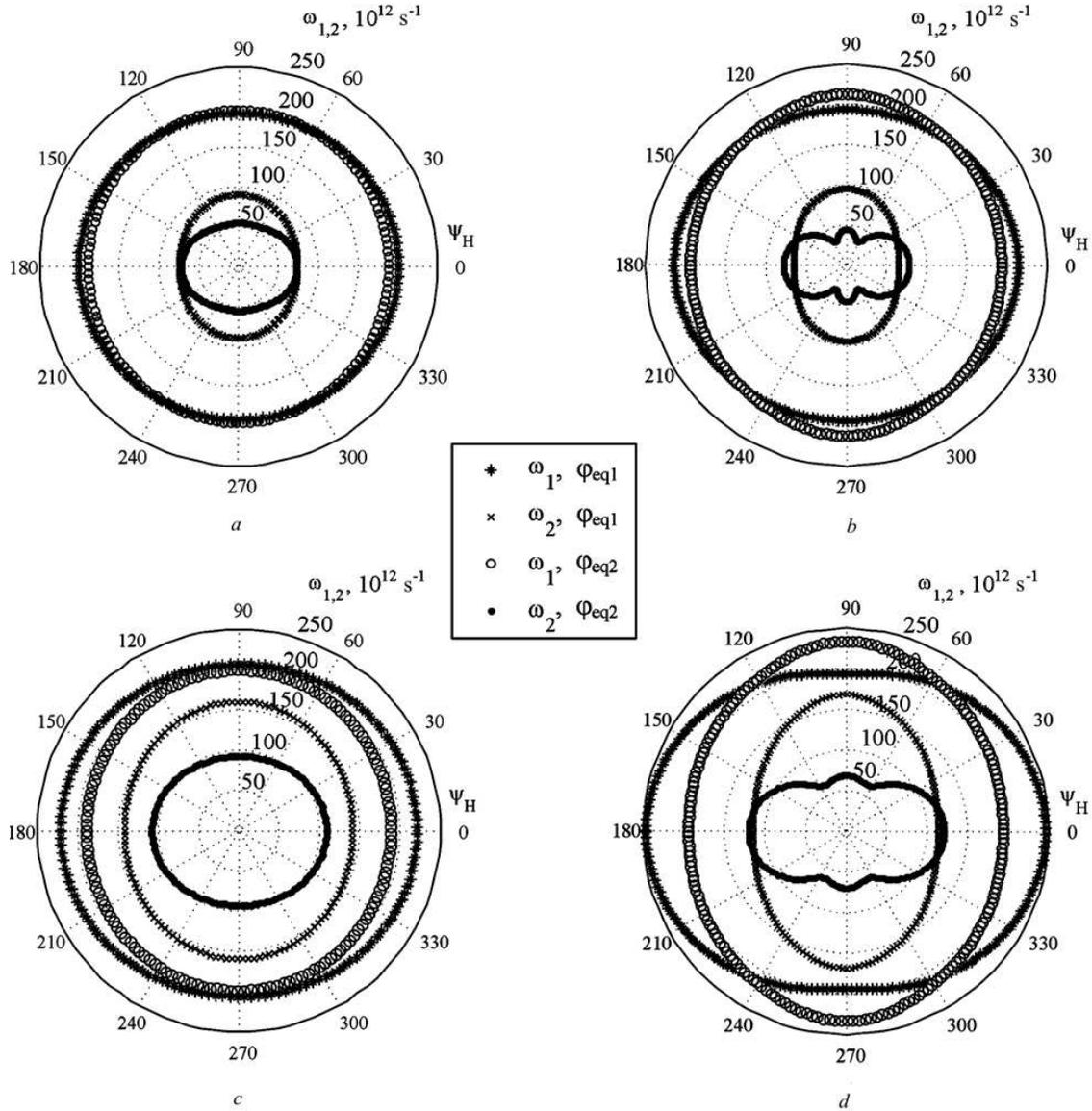}
\vskip-3mm\caption{Dependences of oscillation frequencies
$\omega_{1,2}$ of the AFM vector in a vicinity of two different
equilibrium states $\varphi_{\mathrm{eq1}}$  and
$\varphi_{\mathrm{eq2}}$ (see the insets in Fig.~4) on the magnetic
field direction (the angle $\psi _{\mathrm{H}}$) calculated by
formula (\ref{frequencies}). Panels (\textit{a}) and (\textit{b})
correspond to a specimen with the ratio between its semiaxis lengths
$a/b=2.4$, and panels (\textit{c}) and (\textit{d}) to $a/b=20$. The
magnetic field strength $H=100$ (\textit{a} and \textit{c}), 140
(\textit{b}), and 200~Oe (4). Simulation was carried out for an NiO
AFM particle. The model parameters are quoted in Table
}\label{polar}
\end{figure*}

\section{Parametric Resonance\vspace*{-1mm}}

\textrm{Let us demonstrate that there may emerge a parametric}
resonance in the AFM/PE system in the presence of an ac electric
field. For this purpose, we retain terms down to the $\sigma
_{0}l_{\perp }^{2}$ and $\sigma _{0}l_{\parallel }^{2}$ orders of
smallness in the Lagrange equations (\ref{Lagr_equat}) and present
the mechanical stress that arises in the system owing to the
application of ac electric voltage in the form $\sigma (t)=\sigma
_{0}\cos \omega t$. As a result, we obtain the equations
\[
\ddot l_{\perp} + 2\gamma_{\mathrm{AFM}}\dot
l_{\perp}-\Omega_{\mathrm H} \dot l_{\parallel} + \Omega_{\perp}^2
\left( \!1 + \frac{\sigma_0 \Phi_{\perp}}{\Omega_{\perp}^2} \cos
\omega t\! \right) l_{\perp} = 0; \]\vspace*{-5mm}
\begin{equation}\label{param_res_equat}
\ddot l_{\parallel} + 2\gamma_{\mathrm{AFM}}\dot l_{\parallel} +
\Omega_{\mathrm H} \dot l_{\perp} + \Omega_{\parallel}^2 \left(\! 1
+ \frac {\sigma_0 \Phi_{\parallel}}{\Omega_{\parallel}^2} \cos
\omega t \!\right) l_{\parallel} = 0,
\end{equation}
where $\Phi _{\perp }$ and $\Phi _{\parallel }$ are some functions of
the magnetoelastic coefficients $K_{\perp ,\parallel }^{\mathrm{el}}$,
corrections $\Delta K_{\parallel ,\perp }$, direction $\psi _{\sigma }$ of
the external voltage application, and equilibrium state of the AFM vector, in a
vicinity of which the resonance oscillations (the solutions of
Eq.~(\ref{eq_magn})) are observed.

In the absence of a magnetic field ($\Omega _{\mathrm{H}}=0$), the
oscillation equations (\ref{param_res_equat}) are independent. Each
of the equations acquires a standard form of the Mathieu equation
\cite{Landau}. The parametric resonance with the largest band width
arises at the frequencies
\[
\omega =  2 \Omega_{\perp} \pm \Delta\omega_{\perp}/2; \quad \omega
= 2 \Omega_{\parallel} \pm \Delta\omega_{\parallel}/2. \]%
Here, the first band width for the AFM vector component $l_{\perp}$
equals\vspace*{-2mm}
\[
\Delta\omega_{\perp} = 2 \sqrt{(\sigma_0 \Phi_{\perp} /
2\Omega_{\perp} )^2 - 4\gamma_{\mathrm{AFM}}^2}
\]
and, for component $l_{\parallel}$,\vspace*{-1mm}
\[
\Delta\omega_{\parallel} = 2 \sqrt{(\sigma_0 \Phi_{\parallel}
 / 2\Omega_{\parallel})^2 - 4\gamma_{\mathrm{AFM}}^2}.
 \]

Since $\Omega _{\parallel }\gg \Omega _{\perp }$, it becomes clear
that it is rather a difficult task to excite parametric resonances
in the plane and perpendicularly to it simultaneously. The width
$\Delta \Omega _{\parallel }$ of the first resonance band for
oscillations along the $Oz$ axis is narrower in comparison with that
for in-plane oscillations. This is a consequence of the facts that
the studied antiferromagnetic is of the \textquotedblleft easy
plane\textquotedblright\ type, and the oscillation mode along the
$Oz$ axis is more rigid.

The dependence of the resonance band width on $\Phi _{\perp
,\parallel }$ allows one to regulate the resonance width by
selecting the specimen shape.

The parametric resonance also arises in a vicinity of the frequencies
$\omega =2\Omega _{\perp }/n$ and $\omega =2\Omega _{\parallel }/n$,
where $n=1,2,3,\mbox{...} $\,. The band widths decrease at that, by
following the rules $(\Delta \omega _{\perp })_{n}^{2}\sim (\sigma
_{0}/\Omega _{\perp }^{2})^{2n} $ and $(\Delta \omega _{\parallel
})_{n}^{2}\sim (\sigma _{0}/\Omega _{\parallel }^{2})^{2n}$.

The considered partial case can be realized not only when ${\mathbf{H}}=0$,
but also when $\mathbf{H}$ is oriented perpendicularly to either of the AFM
specimen easy axes (along the $Ox$ or $Oy$ axis). Then, the vector
$\mathbf{L}$ will be in the equilibrium position at either $\varphi _{\mathrm{eq1}}=0$
or $\varphi _{\mathrm{eq}}=\pi /2$, respectively.

In the presence of a magnetic field (${\mathbf{H}}\neq 0$), the complete
system of equations (\ref{param_res_equat}) is to be solved. For this
purpose, let us first change to normal oscillation modes $Q_{1,2}$. The
characteristic frequencies of normal oscillations coincide with the
parameters $\omega _{1,2}$ in expression (\ref{frequencies}). We apply the
Floquet theorem and seek the solutions in the form
\[
Q_1 \sim \exp(i\beta t) \sum_{n} a_n \exp (i \omega nt);
\]\vspace*{-5mm}
\[
Q_2 \sim \exp(i\beta t) \sum_{n} b_n \exp (i \omega nt).
\]

As a result, we obtain a system of equations for the coefficients
$a_{n}$ and $b_{n}$. Using this system to study the parametric
resonance, e.g., in a vicinity of $\omega \approx \omega _{1}$  (for
$n=2$), we obtain the following formula for the resonance band
width:
\[
\Delta \omega_{(n=2)} \approx 2 \sqrt{ \left(\! \frac{\sigma_0^2
\Phi_{\perp}^2}{8 \omega_1^3} \!\right)^{\!\!2} - 4
\gamma_{\mathrm{AFM}}^2}.
\]

Hence, the band width of the parametric resonance for oscillations
of the vector $\mathbf{L}$ in an AFM particle can be controlled by
both changing the geometrical parameters of the particle (this way
allows the magnitudes of resonance frequencies to be adjusted
already at the specimen fabrication stage) and applying external
fields, electric and/or magnetic ones.\vspace*{-2mm}

\section{Discussion and Conclusions}

In this work, the influence of the shape of AFM particles on their resonance
properties has been studied. The proposed model of AFM/PE multiferroic
allows both elastic and magnetic properties of the system to be considered
simultaneously. The combination of a piezoelectric and an antiferromagnet in the same specimen
makes it possible to control the system state by means of electric and
magnetic fields.

It is shown that there emerges an additional uniaxial magnetic anisotropy
in the AFM/PE system, which depends on the specimen shape. This anisotropy
makes one of the antiferromagnet's \textquotedblleft easy\textquotedblright\
axes (they are formed by either the orientation of crystallographic axes or
the application of an external magnetic field) more beneficial
energetically, i.e. it eliminates the degeneration between two equivalent
equilibrium orientations of the AFM vector.

AFM vector oscillations near the stable equilibrium vector
orientation can be excited in a number of cases. In the absence of
external fields, if the antiferromagnet's \textquotedblleft easy\textquotedblright\
axis does not coincide with the easy axes induced by the
specimen shape (for instance, along the ellipse axes), the specimen
deformation can favor the rotation of the AFM vector to a more
energetically beneficial direction. If an ac electric voltage is
additionally applied to the AFM/PE particle, it gives rise to the
emergence of ac mechanical stresses in and, consequently, the
varying deformation of the particle. As a result, the AFM vector
executes small oscillations in the specimen plane (Fig.~3).

For a certain geometry of the system, the mechanical stress cannot
induce resonance oscillations of the AFM vector, namely, when the
stress is applied along one of the \textquotedblleft
easy\textquotedblright\ axes (see Fig.~2,~\textit{a}). The external
magnetic field rotates the vector $\mathbf{L}$ and causes the
appearance of an effective projection of the mechanical stress and,
as a result, the oscillations of the AFM vector owing to the varying
deformation of the specimen (Fig.~4). Moreover, the magnetic field
entangles the \textquotedblleft soft\textquotedblright\ mode of
in-plane oscillations with the \textquotedblleft
rigid\textquotedblright\ mode that is perpendicular to it, so that
the oscillations of the vector $\mathbf{L}$ become elliptically
\mbox{polarized.}\looseness=1

It is of importance that the frequencies of AFM vector resonance
oscillations depend on the specimen shape (for example, on the ratio $a/b$
between the ellipse semiaxis lengths) and on the magnitude and the direction of
an external magnetic field (Fig.~5). This circumstance allows the properties of
an AFM particle to be controlled and predicted already at the particle
fabrication stage; in particular, it is possible to select particle's shape and dimensions, as
well as the corresponding orientation of the crystallographic axes in the
specimen plane, depending on the particle usage specificity.

We have also demonstrated a possibility to excite a parametric
resonance in the AFM/PE system. The band width of the parametric
resonance is governed by both the magnitude of mechanical stress
induced in the specimen and the specimen shape. Hence, by choosing
the shape and the dimensions of a specimen, it is possible to
control the region, where the parametric resonance emerge at
variable \mbox{deformations.}\looseness=1

To summarize, we note that the dependences obtained in this work can be
used for a direct experimental verification of the shape effect in antiferromagnets.

\vskip4mm

\textit{The work was carried out in the framework of the
Target-oriented program for basic researches of the National Academy
of Sciences of Ukraine. It was also partially sponsored by the grant
of the Ministry of Education and Science, Youth and Sport of
Ukraine.}

\subsubsection*{\!\!\!\!\!\!APPENDIX\\
Magnetoelastic Coefficients\\ and the Corresponding Corrections
That Take\\ the Specimen Shape into Account}{\label{app1}}

 {\footnotesize For an elliptic nanofilm ($a>b\gg h$),
the magnetoelastic coefficients look like \cite{GL2007}
\[
K^{\mathrm{el}}_{\parallel}=\frac{h}{b}\frac{[(\lambda^\prime)^2(2-3\nu)+
\lambda_v\lambda^\prime]J_{\parallel}(k)}{4\mu(1-\nu)},
\]
\[
K^{\mathrm{el}}_{\mathrm{is}}=\frac{(\lambda^\prime)^2(3 -
4\nu)}{8\mu(1-\nu)}, \quad
K^{\mathrm{el}}_{\perp}=\frac{h}{b}\frac{2(\lambda^\prime)^2J_{\perp}(k)}{3\mu(1-\nu)},
\]
where $\mu $ is the shear modulus, $\nu $ is Poisson's ratio, and $\lambda
_{v}$ and $\lambda ^{\prime }$ are magnetoelastic moduli. The shape
integrals $J_{\parallel }(k)$ and $J_{\perp }(k)$ depend on the
specimen geometry by means of the parameter $k^{2}=1-b^{2}/a^{2}$ as
follows:
\begin{equation*}
J_{\parallel }(k)=\int\limits_{0}^{\pi /2}\frac{(k^{2}\sin ^{2}\phi
+\cos 2\phi )d\phi }{\sqrt{1-k^{2}\sin ^{2}\phi }},
\end{equation*}
\begin{equation*}
J_{\perp }(k)=\int\limits_{0}^{\pi /2}\frac{(1-8\cos 2\phi
-k^{2}\sin ^{2}\phi +8\cos 2\phi /k^{2})d\phi }{\sqrt{1-k^{2}\sin
^{2}\phi }}.
\end{equation*}%

Elastic deformation of a specimen modifies the anisotropy constants
of the second and fourth orders and contributes to the destressing
energy (\ref{d_w_destr}). The corrections $\Delta K_{\parallel
,\perp }$ depend on the linear sizes of the specimen and the shape
integrals in the following manner \cite{lviv}:
\begin{equation*}
\Delta K_{\parallel }=4\frac{h}{b}\frac{[(\lambda ^{\prime })^{2}(2-3\nu
)+\lambda _{v}\lambda ^{\prime }]}{4\mu (1-\nu
)}(1-k^{2})\frac{dJ_{\parallel }(k)}{d(k^{2})},
\end{equation*}
\begin{equation*}
\Delta K_{\perp }=4\frac{h}{b}\frac{2(\lambda ^{\prime })^{2}}{3\mu (1-\nu
)}(1-k^{2})\frac{dJ_{\perp }(k)}{d(k^{2})}.
\end{equation*}%
 The introduction of those corrections allows one to study how a
change of the specimen shape affects the state of the magnetic subsystem in
an AFM/PE multiferroic.

}

\rezume{%
С.В. Кондович, О.В.~Гомонай, В.М.~Локтєв\vspace*{1mm}}{МАГНІТНА
ДИНАМІКА МУЛЬТИФЕРОЇКА\\ З АНТИФЕРОМАГНІТНИМ
ПРОШАРКОМ\vspace*{0.5mm}} {\rule{0pt}{13pt}Ефекти форми для
магнітних частинок інтенсивно досліджуються, адже форма і розміри
можуть виступати в ролі керуючих параметрів і задавати властивості
зразка вже при його виготовленні. Експерименти дозволяють припустити
існування впливу форми на властивості антиферомагнітних (АФМ)
нанорозмірних зразків, але з теоретичної точки зору цей вплив майже
не розглянуто. В даній роботі запропоновано модель для дослідження
впливу ефектів форми в АФМ частинках на частоті антиферомагнітного
резонансу (АФМР). Методом функцій Лагранжа розраховано спектр
резонансних коливань АФМ вектора для синтетичного мультифероїка
(п'єзоелектрик~+~АФМ). Досліджено вплив форми зразка на частоту АФМР
у присутності зовнішнього магнітного поля. Розглянуто умови, за яких
в магнітній підсистемі виникає: а)~резонанс під дією зовнішньої
примусової сили; б)~параметричний резонанс.}


\vskip2mm
\begin{thebibliography}{99}
\bibitem{FM-1} R.P. Cowburn, J. Phys. D: Appl. Phys. {\bf 33}, R1 (2000).\vspace*{0.5mm}
\bibitem{FM-2} E.Y. Vedmedenko, H.P. Oepen, and J. Kirschner, J.~Magn. Magn.
Mater. {\bf 256}, 237 (2003).\vspace*{0.5mm}
\bibitem{FM-3} Yi Li, Yiran Lu, and W.E. Bailey, J. Appl. Phys. {\bf 113}, 17B506 (2013).\vspace*{0.5mm}
\bibitem{Folven1} E. Folven, T. Tybell, A.~Scholl, A.~Young, S.T.~Retterer,
Y.~Takamura, and J.K.~Grepstad, Nano Lett. \textbf{10}, 4578
(2010).\vspace*{0.5mm}
\bibitem{Folven2} E. Folven, A. Scholl, A.~Young, S.T.~Retterer,
J.E.~Bosch\-ker, T.~Tybell, Y.~Takamura, and J.K.~Grepstad, Phys.
Rev.~B \textbf{84}, 220410 (2011).\vspace*{0.5mm}
\bibitem{Folven3} E.~Folven, A.~Scholl, A.~Young, S.T.~Retterer,
J.E.~Bosch\-ker, T.~Tybell, Y.~Takamura, and J.K.~Grepstad, Nano
Lett. \textbf{12}, 2386 (2012).\vspace*{0.5mm}
\bibitem{Davydov} A.S.~Davydov, \textit{Theory of Molecular Excitons}
(Plenum Press, New York, 1971).\vspace*{0.5mm}
\bibitem{GL2007} H.V. Gomonay and V.M.~Loktev, Phys. Rev.~B \textbf{75},
174439 (2007).\vspace*{0.5mm}
\bibitem{Kornienko} H.V. Gomonay, E.G.~Kornienko, and V.M.~Loktev, Ukr. J.
Phys. \textbf{50}, 816 (2005).\vspace*{0.5mm}
\bibitem{Turov} E.A.~Turov, A.V.~Kolchanov, V.V.~Menshenin, I.F.~Mirsaev,
and V.V.~Nikolaev, \textit{Symmetry and Physical Properties of
Antiferromagnets} (Fizmatlit, Moscow, 2001) (in
Russian).\vspace*{0.5mm}
\bibitem{lviv} S.V. Kondovych and H.V.~Gomonay, Visn. Lviv. Univ. Ser. Fiz.
\textbf{47}, 159 (2012).\vspace*{0.5mm}
\bibitem{NiO-1} P.~de V.~Du Plessis, S.J. van Tonder, and L.~Alberts, J. Phys.~C
\textbf{4}, 1983 (1971).\vspace*{0.5mm}
\bibitem{NiO-2} P.~de V.~Du Plessis, S.J.~van Tonder, and L.~Alberts, J.
Phys.~C \textbf{4}, 2565 (1971).\vspace*{0.5mm}
\bibitem{NiO-3} M.T.~Hutchings and E.J.~Samuelsen, Phys. Rev.~B \textbf{6},
3447 (1972).\vspace*{0.5mm}
\bibitem{Landau} L.D.~Landau and E.M.~Lifshitz, \textit{Mechanics}
(Butterworth-Heinemann, Oxford, 2001).\vspace*{0.5mm}

\begin{flushright}
{\footnotesize Received 29.03.2013.\\ Translated from Ukrainian by
O.I.~Voitenko}
\end{flushright}
\end{thebibliography}
\end{document}